# Theoretical and Experimental Study of LiBH$_4$-LiBr Phase Diagram

Valerio Gulino,[#] Erika Michela Dematteis, Marta Corno, Mauro Palumbo and Marcello Baricco[*]

Department of Chemistry, NIS and INSTM, University of Turin, Via Pietro Giuria 7, 10125 Torino, Italy

[#]Present address: Materials Chemistry and Catalysis, Debye Institute for Nanomaterials Science, Utrecht University, Universiteitweg 99, 3584 CG Utrecht, The Netherlands

*Corresponding author

Marcello Baricco

E-mail address: marcello.baricco@unito.it

Tel.: +39 011 6707569

Fax: +39 0116707856

## Abstract

Because substitutions of $BH_4^-$ anion with $Br^-$ can stabilize the hexagonal structure of the $LiBH_4$ at room temperature, leading to a high Li-ion conductivity, its thermodynamic stability has been investigated in this work. The binary $LiBH_4$-LiBr system has been explored by means of X-ray diffraction and differential scanning calorimetry, combined with an assessment of thermodynamic properties. The monophasic zone of the hexagonal $Li(BH_4)_{1-x}(Br)_x$ solid solution has been defined equal to $0.30 \leq x \leq 0.55$ at 30 °C. Solubility limits have been determined by in-situ X-ray diffraction at various temperatures. For the formation of the h-$Li(BH_4)_{0.6}(Br)_{0.4}$ solid solution, a value of the enthalpy of mixing ($\Delta H_{mix}$) has been determined experimentally equal to -1.0 ± 0.2 kJ/mol,. In addition, the enthalpy of melting has been measured for various compositions. Lattice stabilities of $LiBH_4$ and $LiBr$ have been determined by *ab initio* calculations, using CRYSTAL and VASP codes. Combining results of experiments and theoretical calculations, the $LiBH_4$-LiBr phase diagram has been determined in all composition and temperature range by the CALPHAD method.

## Introduction

Due to its high gravimetric and volumetric density of hydrogen, $LiBH_4$ has been largely studied as a solid-state hydrogen storage material.[1] It shows a polymorphic transition from an orthorhombic structure at room temperature (R$T$), space group (s.g.) *Pnma*, to a hexagonal structure, s.g. *P6₃mc*,[2] above $110 \pm 2$ °C,[3] with an enthalpy change equal to $5.0 \pm 0.9$ kJ/mol.[3] In 2007, Matsuo *et al.*[4] reported a drastic increase of the Li-ion conductivity of $LiBH_4$ above the phase transition temperature, suggesting it as a solid-state electrolyte. Despite the hexagonal polymorph (*h*-$LiBH_4$) shows a remarkable ionic conductivity (~$10^{-3}$ S cm⁻¹ at 120 °C), the orthorhombic room temperature phase (*o*-$LiBH_4$) is much less conductive, showing a Li-ion conductivity of $9.5 \times 10^{-9} \pm 2 \times 10^{-9}$ S cm⁻¹ at 30 °C,[5] making a room temperature battery target unviable.[4]

Different approaches have been used to increase the Li-ion conductivity of $LiBH_4$ at R$T$, such as by mixing it with oxides or by means of nanoconfinement.[6–9] Differently, many studies showed that substitution of $BH_4^-$ anion with halides (*e.g.* I⁻, Br⁻ and Cl⁻) can make the hexagonal structure thermodynamically stable, providing a high ionic conductivity at R$T$.[10–12] For instance, a $LiBH_4$-LiI hexagonal solid solution with 25 mol.% of LiI showed a Li-ion conductivity of about $10^{-4}$ S cm⁻¹ at 30 °C. The *h*-$Li(BH_4)_{1-\alpha}(I)_\alpha$ solid solutions have been reported to be stable at R$T$ in the range of $0.18 \leq \alpha \leq 0.50$.[13] Fast Li-ion conductivity at R$T$ is also observed in *h*-$Li(BH_4)_{1-\alpha}(Br)_\alpha$ hexagonal solid solutions (*e.g.* ~$10^{-5}$ S cm⁻¹ for *h*-$Li(BH_4)_{0.7}(Br)_{0.3}$)[12], although it is reduced as the bromide content increases above $x = 0.29$.[12,14]

A full evaluation of thermodynamic properties of borohydrides and their mixtures is fundamental for a further improvements and insight on complex hydrides, aimed to tailor their properties as hydrogen storage materials and solid-state electrolytes. This goal can be reached with the CALPHAD method, which is based on a parametric description of the Gibbs free energy as a function of temperature and composition, by the combination of *ab initio* calculations and experimental evidences.[15] Starting from experimental data as input, the CALPHAD method allows the assessment of parameters describing the Gibbs free energy of all phases, in order to find the most reliable description of the phase diagram. *Ab initio* calculations are required to establish the Gibbs free energy of compounds with crystal structures that are not stable in the investigated ranges of temperature and pressure. The use of the CALPHAD approach allowed the determination of different thermodynamic properties of borohydrides, *e.g.* isobaric heat capacity, $C_p$,[3] and the definition of phase diagrams as a function of temperature and composition.[16,17]

For the LiBH₄-LiBr phase diagram, only few experimental data and no thermodynamic assessment are present in the literature. Recently, the hexagonal solid solution $h$-Li(BH₄)$_{1-x}$(Br)$_x$ has been demonstrated to be stable at R$T$ in the range $0.29 \leq x \leq 0.50$,[12,14] while a small solubilisation of the LiBr into o-LiBH₄ has been reported (*i.e.* $o$-Li(BH₄)$_{1-x}$(Br)$_x$ where x is $\leq 0.09$)[14]. In addition, the LiBH₄ seems to be insoluble in the cubic LiBr at R$T$.[12,14] Rude *et al.*[18] reported a temperature of melting of 377.9 °C for the $h$-Li(BH₄)$_{0.5}$(Br)$_{0.5}$ solid solution. Therefore, in the present study, we explore the LiBH₄-LiBr system, combining experimental and theoretical approaches, in order to determine its thermodynamics and phase diagram. Literature, experimental and *ab initio* data have been evaluated for an assessment of the system thermodynamics using the CALPHAD approach. The assessment allowed establishing phase stabilities and limits of solubility in the full composition range and in a wide temperature range, *i.e.* from R$T$ up to the liquid phase.

## Experimental

### *Synthesis*

All manipulations were performed in an argon-filled glovebox (MBraun Lab Star Glove Box supplied with pure 5.5 grade Argon, <1 ppm O₂, <1 ppm H₂O). LiBH₄ (purity >95% from Sigma-Aldrich) and LiBr (purity >99% from Sigma-Aldrich), were mixed in different ratios and by different methods, as reported in **Table 1**. LiBr were previously dried by heating at 120 °C under dynamic vacuum, in order to avoid the presence of the hydrated LiBr·H₂O phase.[18]

| Name | Composition (molar fraction) | | Synthesis |
|------|------|------|------|
| | LiBH$_4$ | LiBr | |
| **s1** | 0.4 | 0.6 | BM + AN |
| **s2** | 0.5 | 0.5 | BM + AN |
| **s3** | 0.6 | 0.4 | BM + AN |
| **s4** | 0.7 | 0.3 | BM + AN |
| **s5** | 0.6 | 0.4 | Hand mixed |

**Table 1.** *Composition and synthesis conditions of the samples prepared. BM = ball milling for 1.5 hours, AN = annealing for 2 hours at 250 °C.*

A Fritsch Pulverisette 6 planetary mill was used to ball mill the starting materials in 80 mL tungsten carbide vials, with tungsten carbide balls (10 mm outside diameter). The balls-to-sample mass ratio used was 30:1. The mechanochemical treatment (BM) was performed for 1.5 hours under argon atmosphere at 350 r.p.m. for periods of 10 min of milling, separated by 2 min breaks. In order to reach the equilibrium conditions, samples were annealed (AN) at 250 °C for 2 hours in a quartz tube under static vacuum, with a heating/cooling rate of 5 °C/min. In order to obtain information on the enthalpy of mixing, sample **s5** has been simply hand mixed in a mortar at R$T$.

*Characterization*

*Differential Scanning Calorimetric (DSC)*

A high-pressure 204 Netzsch DSC was used to analyse the thermal behaviour of samples. The instrument is placed inside an Ar-filled glove box to ensure sample handling under inert atmosphere. Approximately 5−10 mg of the sample were loaded into closed aluminium crucibles with a lid. Samples were heated and cooled in the desired temperature range at 5-20 °C/min under 2/15 bars of H$_2$. The instrument was calibrated for temperature and heat flow using the melting temperature and enthalpy of high purity standards (Bi, In, Sn, Zn). The same crucible, heating rate and H$_2$ pressure have been used for measurements and calibrations.

***Attenuated Total Reflection Infrared Spectroscopy (ATR-IR)***

Infrared spectra were collected in ATR-IR mode with a Bruker Alpha-P spectrometer, equipped with a diamond crystal. The instrument is placed inside a nitrogen filled glove box. All spectra were recorded in the 5000−400 cm$^{-1}$ range with a resolution of 2 cm$^{-1}$, and are reported as the average of 50 scans.

***Powder X-ray diffraction (PXD)***

Samples in powder form were characterised by PXD at R$T$ (*ex situ*) using a Panalytical X-pert Pro MPD (Cu K$_{\alpha1}$ = 1.54059 Å, K$_{\alpha2}$ = 1.54446 Å) in Debye-Scherer geometry. Patterns were collected in the 2θ range (from 10° to 70°), with a time step of 60 s, for a total of about 30 min per scan. 0.5 mm glass capillaries were used as sample holder and they were filled and sealed under Ar atmosphere.

***In situ time-resolved synchrotron radiation powder X-ray diffraction (SR-PXD)***

SR-PXD *in situ* measurements were performed at the diffraction beamline P02, in the Petra III storage ring of DESY (Hamburg, Germany). Few milligrams of sample were packed in a single crystal sapphire capillary (inner diameter 0.6 mm). The sample was heated from R$T$ up 370 °C and then cooled down at 5 °C/min Ar atmosphere. The beamline provides a monochromatic X-ray beam (λ = 0.207157 Å) and it is equipped with a PerkinElmer XRD 1621 plate detector (pixel size 200 μm × 200 μm, array 2048 × 2048 pixels). The wavelength and the detector geometry were calibrated using a LaB₆ external standard. The diffraction images, collected every 15 seconds, were integrated with the software Fit2D.

***Rietveld Analysis***

The Rietveld refinement of diffraction patterns has been performed using the MAUD (Materials Analysis Using Diffraction) software.[19] The instrumental function was determined using pure Si, for the *ex situ* measurements and LaB₆ for the *in situ* one. The peak broadening was described using the Caglioti formula,[19] and the peak shape was fitted with a pseudo-Voigt function. Parameters were also refined to consider possible instrument misalignments. Reliability parameters R$_{wp}$, R$_{exp}$ and χ$^2$, were used to evaluate the quality of the fitted patterns with selected structural and microstructural (*i.e.*

crystallites size and micro-strain) parameters. The background was described through a polynomial function with 3 or 4 parameters. The following sequence was applied for the refinement of parameters: (1) scale factor (2) background parameters (3) lattice constants (4) crystallites size (5) micro-strain. In some cases, also the occupancy and the position of the *2b* site in the hexagonal structure were refined.

## Modelling

### *Ab initio*

### *CRYSTAL*

The adopted level of theory for the computational study is in the framework of density functional theory (DFT) with the generalized gradient approximation (GGA) PBE functional.[20] The calculations were performed using the periodic quantum-mechanical software CRYSTAL17,[21,22] which utilizes localized Gaussian functions to describe electrons. In details: lithium cation was described with a 5–11G(d) basis set ($\alpha_{sp}$ = 0.479 bohr$^{-2}$ for the most diffuse shell exponent and $\alpha_{pol}$ = 0.600 bohr$^{-2}$ for polarization), boron with a 6–21G(d) ($\alpha_{sp}$ = 0.124 bohr$^{-2}$ for the most diffuse shell exponent and $\alpha_{pol}$ = 0.800 bohr$^{-2}$ for polarization), hydrogen with a 31G(p) ($\alpha_{sp}$ = 0.1613 bohr$^{-2}$ for the most diffuse shell exponent and $\alpha_{pol}$ = 1.1 bohr$^{-2}$ for polarization) and bromide with a large-core pseudopotential basis set ($\alpha_{sp}$ = 0.107 bohr$^{-2}$ for the most diffuse shell exponent).[23] Geometry optimization and phonons at $\Gamma$ point in the harmonic approximation on the optimized geometry were computed to derive the thermodynamic functions by diagonalizing the associated mass-weighted Hessian matrix (for details on the computational procedure see references).[24,25] Enthalpy data were obtained by computing the electronic energy, inclusive of the zero-point energy correction (ZPE), and the thermal factor at T =25 °C.

### *VASP*

Ground state energies at 0 K [-273 °C] were also computed using DFT as implemented in the Vienna *Ab initio* Simulation Package (VASP) with plane-wave basis-sets.[26–28] The calculations employed the GGA of Perdew *et al.*[20] (PBE) and a cut-off energy of 800 eV. The valence electrons were represented by projector augmented wave (PAW) pseudopotentials, and the k-point meshes were created using the Monkhorst–Pack scheme.[29] The density of the mesh was chosen to guarantee a numerical accuracy of <1 meV/at. In the calculations, the PAW pseudopotentials provided within the VASP package for all elements (Li_sv: 1s2s2p, H: 1s, B: 2s2p, Cl, Br, I: 2s2p) were employed. The ground state energy (at T = 0 K [-273 °C]) was determined by structural relaxation using the Methfessel–Paxton method[30] of order 1. A final step using the tetrahedron method with Blochl corrections[31] was performed to obtain accurate energy values.

To verify the vibrational contribution to energy differences, phonon calculations were carried out using the PBE functional. The vibrational contribution to the free energy was calculated using Density Functional Perturbation Theory as implemented in VASP (IBRION = 8) to determine the dynamical matrix of the system. The Phonopy code[32] was then used to extract the force constants matrix, to calculated phonon dispersions, density of states (DOS) and thermodynamic properties. 2x2x2 supercells were used for the calculations of the dynamical matrix.

*Quantum Espresso*

To further investigate *ab initio* results at 0 K [-273 °C], a limited number of calculations were carried out with the Quantum Espresso (QE) package.[33–35] Similarly to VASP calculations, we used a Methfessel-Paxton smearing scheme with 0.02 Ry width and a final step with the tetrahedron method with Blochl corrections for accurate energy values. The integration over the Brillouin Zone (BZ) was performed employing a Monkhorst-Pack k-points mesh. The density of the mesh was chosen to guarantee a numerical accuracy of < 1 meV/at. The cut-off energy was set to 100 Ry. We employed both ultrasoft (US) and Projected Augmented Wave (PAW) pseudopotentials from the pslibrary 1.0[36] and specifically the following pseudopotentials: Li.pbe-s-kjpaw_psl.1.0.0.UPF, Br.pbe-n-kjpaw_psl.1.0.0.UPF, Li.pbe-sl-rrkjus_psl.1.0.0.UPF, Br.pbe-n-rrkjus_psl.1.0.0.UPF.

*CALPHAD*

The Gibbs free energy of single phases was described according to the CALPHAD approach:[15]

$$^{\varphi}G = {}^{\varphi}G^{ref} - TS^{id} + {}^{\varphi}G^{exc} \qquad (1)$$

$$^{\varphi}G^{ref} = x^{\varphi}G(LiBH_4) + (1-x)^{\varphi}G(LiBr) \qquad (2)$$

$$S^{id} = -R[xln(x) + (1-x)ln(1-x)] \qquad (3)$$

where $\varphi$ is the considered phase (*i.e.* CUB: cubic, ORTHO: orthorhombic, HEX: hexagonal, LIQ: liquid), $x$ is the molar fraction of LiBH₄, $T$ is the temperature and $G^{ref}$, $-TS^{id}$ and $G^{exc}$ are the reference, ideal and the excess contributions to the Gibbs energy, respectively. Excess Gibbs energy was modelled with Redlich-Kister expansion series[37] truncated to the first contribution, since the agreement with thermodynamic data was satisfactory:

$$^{\varphi}G^{exc} = x(1-x)(a+bT) \qquad (4)$$

where $a$ and $b$ are optimized parameters. When $b$=0, $a$ corresponds to the interaction parameter, $^{\varphi}\Omega$, in the regular solution model.

Starting from the enthalpy difference between the stable and the metastable structures, as obtained from *ab initio* calculations, thermodynamic functions for missing end-members (*i.e.* ORTHO-LiBr, HEX-LiBr and CUB- LiBH₄) were evaluated adding the assessed values to the Gibbs energy of the stable phases.

## Results and Discussion

*Structural characterization*

The hexagonal *h*-Li(BH₄)₁₋ₓ(Br)ₓ solid solution has been reported to be stable in the range $0.29 \leq x \leq 0.50$[12,14] and a small solubilisation of the LiBr into o-LiBH4 has been reported (*i.e.* o-Li(BH₄)₁₋ₓ(Br)ₓ with $x \leq 0.09$)[14]. In addition, the LiBH₄ seems to be insoluble in the cubic LiBr at R*T*.[12,14] In order to confirm the limit of solubility of LiBr into Li(BH₄), samples **s1** and **s2** (see **Table 1**) have been analysed by PXD, and results are shown in **Figure S1.** The formation of a single hexagonal solid solution is confirmed for sample **s2** (*i.e.* Li(BH₄)₀.₅(Br)₀.₅), whereas a two-phase mixture has been observed for sample **s1**. To define the composition limits of the bromide rich biphasic (*i.e.,* hexagonal and cubic) zone as a function of temperature, *in situ* SR-PXD measurement was performed on sample **s1** (**Figure 1**).

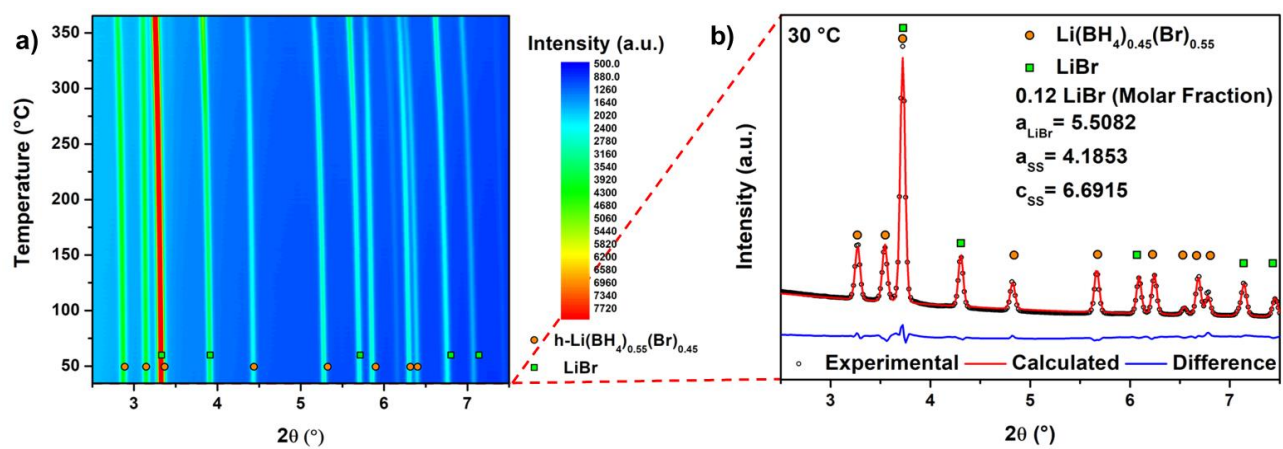

**Figure 1.** *a) in situ SR-PXD data for sample **s1** heated from RT to 370 °C (heating rate 5 °C/min) in Ar atmosphere. Temperatures have been calibrated using LiBr as internal standard. b) Rietveld refinement of sample **s1** at 30 °C ($R_{wp}$ 3.23 %). Values of lattice constants for LiBr and hexagonal solid solution (SS) are reported in Å.*

**Figure 1b** show the SR-PXD pattern of sample **s1** at 30 °C collected after the synthesis (*i.e.* at the beginning of the measurement), where two different phases are observed: $h$-$Li(BH_4)_{1-x}(Br)_x$ hexagonal solid solution and cubic LiBr ($Fm\overline{3}m$). For the Rietveld refinement, the hexagonal solid solution has been considered isostructural to the hexagonal polymorph of $LiBH_4$ ($P6_3mc$),[2] which was used as initial structural mode. It is worth nothing that the strong X-ray scattering of the $Br^-$ gives an unambiguous and robust identification of the position and occupancy of the anion. Initially, the $Br^-$ has been placed in the same *2b* site of the $BH_4^-$ anion in pure $LiBH_4$ ($x = 0.3333$, $y = 0.6667$, $z = 0.553$)[2], but after the refinement a small increase of the $z$ coordinate was obtained. A change in the $z$ position for the anion in the *2b* site was already detected by Cascallana-Matias *et al.*[14], confirming obtained results of the refinement. The bromide and boron thermal displacement parameters ($U_{iso}$) were bounded to the same value.[14] The Rietveld refinement output parameters are reported in **Table S1**.

It is possible to exclude any solubilisation of $BH_4^-$ inside the cubic structure of LiBr, since the cell parameter of the cubic phase ($a = 5.5082$ Å) is equal to that of pure LiBr, as obtained by a Rietveld refinement of the starting material. For this reason, the temperature has been calibrated considering LiBr as internal standard, using the volumetric expansion coefficients reported by Rapp *et al.*[38].

The lattice parameters of the hexagonal solid solution at 30 °C, obtained from the Rietveld refinement (**Figure 1b**), are $a = 4.1853$ Å and c = 6.6915 Å, in agreement with the lattice parameters obtained from the *ex situ* pattern at R*T* (*i.e.* $a = 4.1861$ Å, c = 6.6940 Å, **Figure S1**). In addition, a $Li(BH_4)_{0.45}(Br)_{0.55}$ composition has been obtained from the *2b* site occupancy, which is present with a molar phase fraction equal to 0.88, together with pure LiBr.

After the refinement, a molar balance has been applied, according to:

$$f\ Li(x\ BH_4 \cdot (1-x)\ Br) + (1-f)\ LiBr = 1 \qquad (5)$$

where $f$ and $(1-f)$ are the molar phase fractions of the hexagonal solid solution $Li(BH_4)_{1-x}(Br)_x$ and of LiBr, respectively; $x$ and $(1-x)$ are the molar fractions of $BH_4^-$ and $Br^-$, respectively, in the hexagonal structure, *i.e.* the occupancy of the *2b* site. Resolving the molar balance, considering a $Li(BH_4)_{0.45}(Br)_{0.55}$ composition for the hexagonal solid solution and the obtained phase molar fractions, results confirmed the output of the Rietveld refinement, suggesting that the monophasic zone can be redefined slightly higher than the x ≤ 0.50 value reported by Cascallana-Matias *et al.*[14].

In order to obtain the structural information and composition of the $h-Li(BH_4)_{1-x}(Br)_x$ as a function of temperature, Rietveld refinement has been performed on SR-PXD patterns collected at different temperatures and results are reported in **Figures S2-S7**. The position of the $Br^-$ anion after the refinement remained $x = 0.3333$, $y = 0.6667$ and $z = 0.609$, with an occupancy equal to 0.55, throughout the investigated temperature range. Once the structural parameters were defined by Rietveld refinement, it was possible to evaluate the lattice constants and the unit cell volume of the two phases (**Figure 2a**) and the molar phase fractions (**Figure 2b**) as a function of temperature.

During heating up to 230 °C, *in situ* SR-PXD (**Figure 1**) shows that no further solubilisation of Br⁻ into *h*-Li(BH₄)₁₋ₓ(Br)ₓ solid solution occurs. In fact, the molar phase fractions remain nearly constant (**Figure 2b**). Above 230 °C, the LiBr molar fraction increases (**Figure 2b**), indicating that a possible decomposition of the hexagonal solid solution occurred, as expected because of the presence of BH₄⁻ anion in the structure. In fact, pure LiBH₄ is expected to decompose at about 230 °C under a partial pressure of H₂ close to $10^3$ Pa,[3] which is compatible with the Ar atmosphere present in the sample holder. During the decomposition, BH₄⁻ likely transforms to volatile products, whereas remaining Li⁺ and Br⁻ combine to form LiBr. This decomposition mechanism is confirmed considering that, in the patterns collected at temperatures higher than 230 °C, the presence of additional crystalline phases, *i.e.* possible decomposition products of the solid solution, have not been detected. The Rietveld refinement results obtained for the last pattern collected after the cooling at about 135 °C (**Figure S8**) shows that the molar fractions of LiBr is nearly the same than that obtained at 370 °C, *i.e.* 0.24, confirming the occurrence of irreversible transformations during heating. These results suggest that data obtained at temperatures higher than 230 °C cannot be used for further analyses, so that only reliable data have been summarized in **Table S1**.

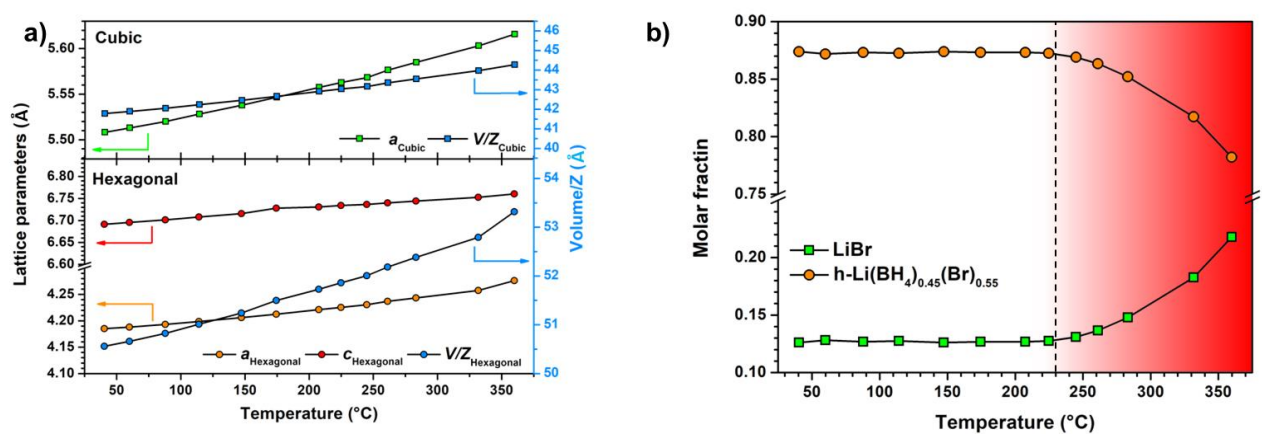

***Figure 2.*** *a) Lattice parameters and unit cell volume (V/Z) as a function of temperature for h-Li(BH₄)₁₋ₓ(Br)ₓ hexagonal solid solution and cubic LiBr. b) Molar phase fraction of h-Li(BH₄)₁₋ₓ(Br)ₓ hexagonal solid solution and cubic LiBr as a function of temperature.*

The volumetric thermal expansion coefficient of the pure hexagonal LiBH$_4$ phase has been reported to be equal to $2.9 \times 10^{-4}$ °C$^{-1}$.[39] From data reported in **Figure 2a** in the temperature range from R$T$ up to 370 °C, we estimated a volumetric thermal expansion coefficient for the Li(BH$_4$)$_{0.45}$(Br)$_{0.55}$ hexagonal solid solution equal to $1.9 \times 10^{-4}$ °C$^{-1}$, indicating that the presence of the Br$^-$ slightly reduces the volumetric thermal expansion coefficient of the hexagonal phase. Arnbjerg *et al.*[40] also observed, for h-Li(BH$_4$)$_{1-x}$(Cl)$_x$ solid solutions, a reduction of the volumetric thermal expansion coefficient, reported to be equal to $1.33 \times 10^{-4}$ °C$^{-1}$ and $1.99 \times 10^{-4}$ °C$^{-1}$ for Li(BH$_4$)$_{0.71}$(Cl)$_{0.29}$ and Li(BH$_4$)$_{0.58}$(Cl)$_{0.42}$, respectively. The reduction of volumetric thermal expansion of LiBH$_4$-LiBr and LiBH$_4$-LiCl solid solutions with respect to the pure LiBH$_4$ compound, can be explained considering the lower ionic radius of the Br$^-$ and Cl$^-$ with respect to BH$_4^-$, which is responsible for an increase of the bond strength that promotes the formation of a solid solution.[41]

Since the bromide-rich biphasic zone has been defined as a function of temperature, samples *s3* and *s4* (see **Table 1**) have been synthetized in order to verify the hexagonal solid solution monophasic zone. **Figure S9a** shows PXD patterns of the samples **s3** and **s4**, together with **s2**, after the synthesis (ball milling followed by a thermal treatment). In all the patterns, only the high temperature hexagonal phase of LiBH$_4$ is present, confirming that it is stabilized at room temperature for all mixtures. **Figure S10** shows lattice parameters and the unit cell volume ($V/Z$) as a function of the molar fraction of the bromide in the Li(BH$_4$)$_{1-x}$(Br)$_x$ hexagonal solid solutions. These values have been calculated by Rietveld refinement of the PXD of samples **s2**, **s3** and **s4**, and are reported together with those obtained from the R$T$ SR-PXD pattern of sample **s1**. Results are in good agreement with previously reported data.[14] The lattice parameters and the unit cell volume linearly decrease increasing the bromide concentration, according to the difference in the BH$_4^-$ and Br$^-$ anion dimensions. By a linear fit, it was possible to define three equations describing the lattice parameters and cell volume as a function of the bromide content inside the hexagonal solid solution. From the value of the volumetric thermal expansion coefficient ($2.9 \times 10^{-4}$ °C$^{-1}$)[39], a $V/Z$ value of 53.3 Å$^3$ has been extrapolated for metastable $h$-LiBH$_4$ at R$T$, which corresponds to the intercept of the linear fit, further confirming the reduction of the unit cell volume due to the formation of the solid solution.

To study the changes in the vibrational properties of lithium borohydride, due to the stabilization of the hexagonal phase by halide additions, IR-ATR spectroscopy was performed on samples **s2**, **s3** and **s4** and results are shown in **Figure S9b**, where the IR-ATR spectrum for pure $o$-LiBH$_4$ is also reported for comparison. As already reported,[18,42–44] the $o$-LiBH$_4$ spectrum shows two main absorption bands, *i.e.* in the 2400–2000 cm$^{-1}$ region and 1600–800 cm$^{-1}$ region, corresponding, respectively, to the B–H stretching and bending vibrational modes. The changes in the spectra of hexagonal solid solutions, in both adsorption bands of the B–H, has been reported to be related to the change of BH$_4^-$ site symmetry.[18] In fact, similar behaviour has been observed for the hexagonal phase of LiBH$_4$ stabilized at R$T$ by Cl$^{-}$[41], Br$^{-}$[18] and I$^{-}$[44] substitutions.

### *Enthalpy of Mixing*

To assess the LiBH$_4$-LiBr phase diagram, a value of the enthalpy of mixing for hexagonal solid solutions is needed. For this reason, sample **s5** (0.4LiBr-0.6LiBH$_4$ molar fractions) has been prepared by hand mixing (HM) in an agate mortar for about 5 min, in order to intimately put in contact the two components, but avoiding the formation of the hexagonal solid solutions. In fact, Rude *et al.*[18] reported that a small amount of $h$-Li(BH$_4$)$_{1-x}$(Br)$_x$ was already stabilized at R$T$, after 15 min of HM. Samples **s5** was later analysed using DSC under 2 bar of H$_2$. The program temperature provided: a) a 2.5 h isotherm at 60 °C, in order to; b) a fast heating ramp (20 °C/min), in order to limit the temperature range in which the thermal activated mixing process could occurs, and c) a 2.5 h isotherm, at the maximum temperature reached during the heating ramp (250 ÷ 350 °C), to ensure that the possible activated thermal process could be completed and to equilibrate again the DSC signal. The same temperature program was repeated twice on the same sample, in order to have a DSC ramp to be used as baseline for the signal integration.

In all the calorimetric analyses of sample **s5** (**Figure S11**), during the heating ramp, the endothermic peak due to the phase transition of the LiBH$_4$ at 110 °C was detected. At higher temperatures, a broad exothermic DSC signal was observed, which has been associated to the formation of the hexagonal solid solution. So, from its integration, a value of the enthalpy of mixing can be obtained, as described below. It is worth noting that, during the second ramp, the endothermic peak due to the phase transition of the LiBH$_4$ is not present anymore, suggesting that the formation of the $h$-Li(BH$_4$)$_{1-x}$(Br)$_x$, is completed after the first ramp. The cooling ramps, from the high temperature isotherms down to 60 °C, were also collected, but they have not been reported, since no thermal events were detected, confirming the formation of the solid solution during the first ramp. The measurement performed with a maximum temperature up to 350 °C (**Figure S11d**) shows that an additional endothermic peak is present at high temperatures. This can indicate that, in this temperature range, the sample starts to melt. The higher stability of the sample obtained in the DSC measurement on sample **s5** (**Figure S11d**) with respect to that observed in the SR-PXD measurement in sample **s1** (**Figure 1**.) can be explained considering that the overpressure of H$_2$ (2 bar) present during calorimetric analysis suppress possible decompositions.[3]

**Figure S12** shows that the difference between the first and the second DSC ramp is composed by an endothermic signal, followed by an exothermic signal at higher temperatures. The peak and onset temperatures of these peaks, as well as the corresponding enthalpies changes, are reported in **Table 2** for the different calorimetric analysis.

The endothermic peak ($\Delta H_{Endo}$) is characterised by two components (**Figure S12**). The main contribution ($\Delta H_{Trs}$) can be assigned to the transition for the orthorhombic to the hexagonal phase of the pure LiBH$_4$, which is present in the hand mixed sample **s5**. The observed signal has been normalized for its content ($\Delta H_{Trs/Normalized}$) and results indicate that, before and during the phase transition, no significant solubilisation occurs. In fact, these values are comparable to that of the pure LiBH$_4$ (*i.e.* $\Delta H_{Trs}$ = 4.89 kJ/mol, see **Figure S13**). The second contribution to the endothermic event is observed before the phase transition, in the temperature range from 60 °C to 120 °C. This signal ($\Delta H_{Cp}$) can be explained considering that the molar heat capacity (C$_p$) of the orthorhombic phase is higher with respect to that of the hexagonal phase.[45] In fact, during the first heating, LiBH$_4$ is still in the orthorhombic phase, while during the second one, LiBH$_4$ have been stabilized in its hexagonal phase, forming the solid solution.

**Figure 3** collects the differences between the first and second DSC signals obtained for different maximum temperatures and results of the integration of the exothermic peak ($\Delta H_{Exo}$) are reported in **Table 2**.

| Max T | T$_{Peak}$ | T$_{Onset}$ | $\Delta H_{Endo}$ | $\Delta H_{Trs}$ | $\Delta H_{Trs/Normalized}$ | $\Delta H_{Cp}$ | $\Delta H_{Exo}$ |
|---|---|---|---|---|---|---|---|
| °C | °C | °C | kJ/mol$_{mix}$ | kJ/mol$_{mix}$ | kJ/mol$_{LiBH4}$ | J/mol$_{mix}$ | J/mol$_{mix}$ |
| 250 | 111 | 93 | 3.71 | 3.01 | 5.02 | 744 | -302 |
| 270 | 111 | 94 | 3.52 | 2.83 | 4.72 | 684 | -471 |
| 285 | 111 | 95 | 3.35 | 2.69 | 4.48 | 660 | -636 |

***Table 2.*** *$\Delta H$, peak and onset temperatures collected during the DSC analysis. Max T gives the maximum temperature reached during the measurement; T$_{Peak}$ and T$_{Onset}$ are the peak and onset temperatures, respectively, for the LiBH$_4$ phase transition; $\Delta H_{Endo}$ corresponds to the integration of the of the entire endothermic signal (see also **Figure S12**); $\Delta H_{Trs}$ was obtained integrating from the T$_{Onset}$ the main endothermic peak due to the LiBH$_4$ phase transition; $\Delta H_{Trs/Normalized}$ corresponds to $\Delta H_{Trs}$, normalized for the LiBH$_4$ molar fraction in the sample (kJ/mol$_{LiBH4}$); $\Delta H_{Cp}$ refers to the contribution due to the difference in the heat capacity between the orthorhombic and hexagonal phase of the LiBH$_4$ in the temperature range from 60 to 120 °C and it has been calculated as $\Delta H_{Cp} = \Delta H_{Endo} - \Delta H_{Trs}$; $\Delta H_{Exo}$ corresponds to the integration of the exothermic signal up to the start of the high temperature isotherm.*

Increasing the temperature of the final isotherm, the exothermic peak increases in enthalpy, reaching a maximum value when the isotherm was set at 285 °C. Regarding the calorimetric analysis in which the isotherm was 250 °C and 270 °C, the exothermic event closes at the end of the heating ramp, while for the isotherm at 285 °C, the exothermic event proceeds also during the isotherm. The different values of $\Delta H_{Exo}$ observed at different temperatures (**Table 2**), indicate that the mixing reaction is not completed during the heating step of the DSC measurements. For the DSC measurement at 285 °C, the integration of the detectable signal after 8 minutes of isotherm (**Figure S12c**) provided a value for $\Delta H_{Exo}$ equal to -792 J/mol$_{mix}$. This result suggests that the kinetics of mixing is very sensitive of holding temperature. In fact, during the isotherms at 250 °C and 270 °C, a rather long time is necessary to reach the complete mixing (*i.e.* all the isothermal annealing at the maximum temperature), hindering the measurement of corresponding heat by DSC. On the contrary, at 285 °C, the mixing reaction is faster, allowing the measurement of a fraction of the heat of mixing also in isothermal conditions.

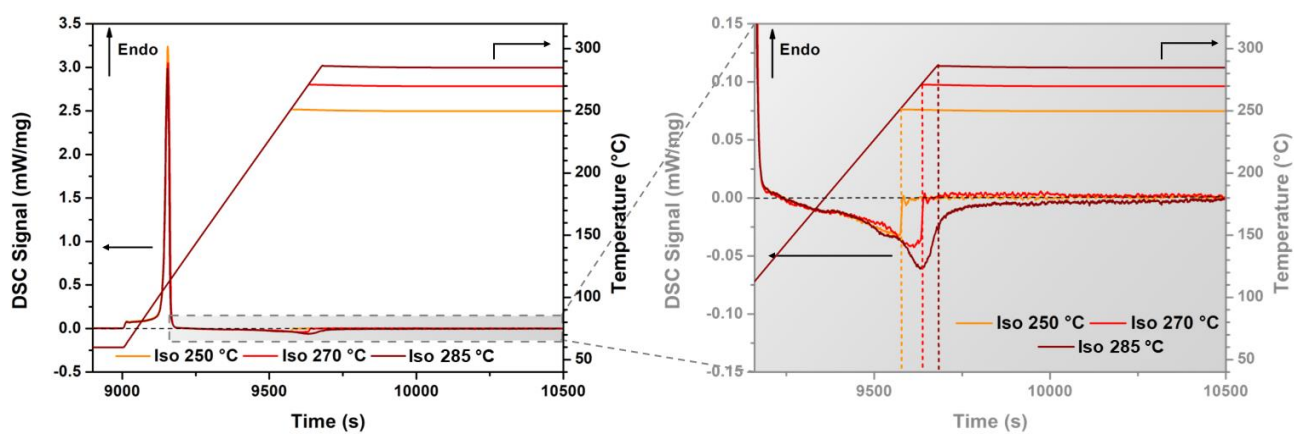

***Figure 3.*** *Overlapping of the DSC signals after the subtraction of the second cycle from the first cycle for the calorimetric analysis.*

In order to define the value of the enthalpy of mixing ($\Delta H_{Mix}$), the reaction was followed by PXD in separate experiments (**Figure S14**). Increasing the maximum temperature reached during the heating ramp, but avoiding the isothermal step, the amount of LiBr and *o*-LiBH$_4$ continuously decreases, while the amount of the hexagonal solid solution increases. In all cases, the composition of the *h*-Li(BH$_4$)$_{1-x}$(Br)$_x$, solid solution remains almost constant ($0.50 \leq x \leq 0.51$) during the reaction. Since the **s5** sample was annealed with the same temperature program used during the first ramp of the DSC analysis, it is possible to correlate the reaction coordinate obtained by Rietveld refinement of PXD patterns (**Figure S14**) with the $\Delta H_{Exo}$ values reported in **Table 2**, as shown in **Figure 4**. For the experiment performed at 285 °C, both values obtained at the end of the heating ramp and after 8 minutes of isothermal holding have been considered.

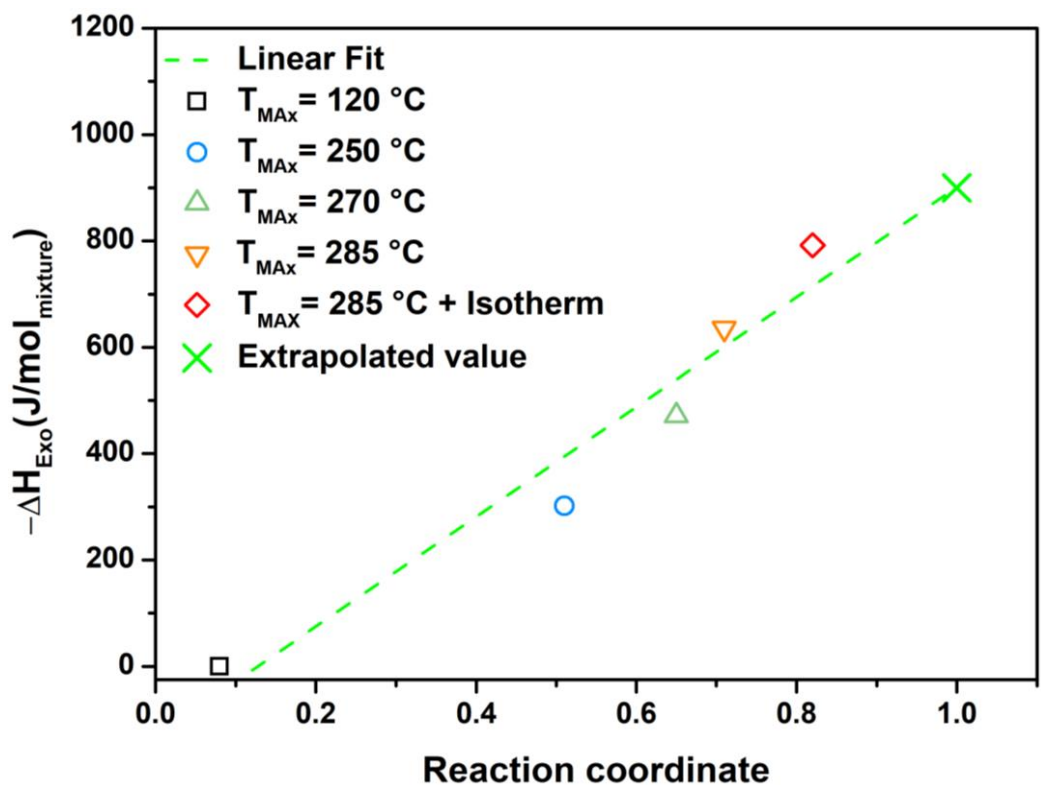

***Figure 4.*** *$\Delta H_{Exo}$ measured by DSC as a function of the reaction coordinate for the formation of the LiBH₄-LiBr hexagonal solid solution in **s5**.*

The reaction coordinate in **Figure 4** corresponds to the fraction of the LiBr solubilized in the hexagonal solid solution and it has been calculated using the molar fraction of residual cubic LiBr obtained by the Rietveld refinement of PXD patterns. The exothermal peak corresponding to the pattern collected after the heating at 120 °C (the offset temperature of the phase transition peak) has been considered equal to zero. However, a small amount of the hexagonal solid solution was already observed (see **Figure S14**), suggesting that the reaction might be already initiated during the heating up to the LiBH₄ phase transition, but the enthalpy contribution cannot be determined due to sensitivity limitation of the DSC analysis. By a linear fit (95 % of confidence, equation determined: y = 1032x -132) of the data collected, the value of $\Delta H_{Exo}$ corresponding to a complete formation reaction of *h-* $Li(BH_4)_{0.6}(Br)_{0.4}$ solid solution has been estimated equal to -900 J/mol (**Figure 4**). Taking into account the intercept of the linear fitting, which represents the missed contribution to the heat of mixing reaction, the measured values of $\Delta H_{Exo}$ should be corrected by adding 132 $J/mol_{mixture}$. As a conclusion, using a confidence interval of 95 %, the value for the enthalpy of mixing for sample **s5** can be considered with an error of $\pm$ 160 J/mol, *i.e.* equal to -1.0 $\pm$0.2 kJ/mol.

### *Ab initio*

Lattice stabilities for $LiBH_4$ have been taken from the literature.[16] In order to determine the relative stability of metastable LiBr structures and to support the assessment of related CALPHAD end-members energies, different theoretical approaches and computer codes (VASP, CRYSTAL and Quantum Espresso) were used. Energy results are reported in **Table 3** as differences between the metastable structures (HEX and ORTHO) and the stable one (CUB). Metastable structures were obtained for both the orthorhombic and hexagonal phases as the full substitution of $BH_4^-$ with bromine (end-members). Results at 0 K [-273 °C] include the electronic energy (EL) and the Zero Point Energy (ZPE) contributions. In some cases, the phonon contribution at 298 K [25 °C] ($\Delta F_{vib}$) has been also included.

It is noteworthy that the calculated values of the energy differences are rather small for all structures. It can also be appreciated that the differences among results obtained with different computer codes are limited, in particular between VASP and QE. Results are almost the same when ultrasoft and PAW pseudopotentials are used with QE. The present results are similar to those reported by Čančarević *et al.*[46] for the energy difference between hexagonal and cubic structures of LiBr. According to this investigation, the most stable structure is either the cubic or hexagonal one, depending on the computational approaches and settings used (including DFT with B3LYP functional and Hartree-Fock).

| Structures | Approach | Energy contributions | Temperature (K [°C]) | ΔE (eV) | ΔE (kJ/mol) |
|---|---|---|---|---|---|
| **HEX-CUB** | CRYSTAL | EL | 0 [-273 °C] | 0.0300 | 2.840 |
| | CRYSTAL | EL + ZPE | 0 [-273 °C] | 0.0520 | 4.940 |
| | CRYSTAL | EL + $\Delta F_{vib}$ | 298 [25 °C] | 0.0740 | 7.120 |
| | VASP | EL | 0 [-273 °C] | -0.0560 | -5.800 |
| | QE US | EL | 0 [-273 °C] | -0.0580 | -5.740 |
| | QE PAW | EL | 0 [-273 °C] | -0.0580 | -5.700 |
| | CALPHAD | - | 298 [25 °C] | 0.0136 | 1.325 |
| **ORTHO-CUB** | CRYSTAL | EL | 0 [-273 °C] | -0.0046 | -0.440 |
| | CRYSTAL | ZPE | 0 [-273 °C] | 0.0240 | 2.300 |
| | CRYSTAL | EL + $\Delta F_{vib}$ | 298 [25 °C] | 0.0420 | 4.060 |
| | VASP | EL | 0 [-273 °C] | -0.0076 | -0.800 |
| | VASP | EL+ZPE | 0 [-273 °C] | -0.0024 | -0.240 |
| | VASP | EL+$\Delta F_{vib}$* | 298 [25 °C] | -0.0070 | -0.680 |
| | CALPHAD | - | 298 [25 °C] | 0.0440 | 4.304 |

***Table 3.** Ab initio results for the LiBr system, compared with optimised values from CALPHAD assessment. Energies are reported as kJ/mol of compound. All calculations were performed with the PBE functional. EL = electronic energy; ZPE = Zero Point Energy; $\Delta F_{vib}$ = phonon contribution at 298 K [25 °C]; US = ultrasoft pseudopotentials; PAW = Projected Augmented Waves pseudopotentials. *some imaginary frequencies were found for the orthorhombic phase.*

The phonon contribution to the electronic energy does not significantly change these differences, which remain rather small. In particular, according to VASP and QE calculations, the hexagonal structure still appears more stable than the cubic one, in contrast to experimental evidence. Other calculations carried out with different functionals show that, in fact, computational settings are particularly relevant for these systems. For example, functionals which include empirical corrections for dispersion interactions (DFT-D2, DFT-D3, etc.)[47,48] have shown a significant effect on the energy differences reported in **Table 3**. Additional calculations with different computational settings to further investigate this point are ongoing and will be presented in a future work.

| Structures | Source | a (Å) | b (Å) | c (Å) |
|---|---|---|---|---|
| CUB | Experimental (TW) | 5.5082 | - | - |
| | Experimental (Ref.[49]) | 5.5000 | - | - |
| | Experimental (Ref.[46]) | 5.5100 | - | - |
| | Experimental (Ref.[14]) | 5.4942 | - | - |
| | VASP | 5.5078 | - | - |
| | QE (US) | 5.5078 | - | - |
| | QE (PAW) | 5.5078 | - | - |
| | CRYSTAL | 5.5232 | - | - |
| HEX | VASP | 4.1933 | - | 6.7341 |
| | QE (US) | 4.1933 | - | 6.7352 |
| | CRYSTAL | 4.2065 | - | 6.6820 |
| ORTHO | VASP | 4.4766 | 7.7099 | 5.7080 |
| | CRYSTAL | 4.4751 | 7.7565 | 5.7260 |

***Table 4.*** *Ab initio LiBr lattice parameters calculated with different codes for different crystal structures, compared with experimental values determined in this work (TW) and taken from the literature. All calculations were performed with the PBE functional. US = ultrasoft pseudopotentials; PAW = Projected Augmented Waves pseudopotentials.*

The lattice parameters calculated with different approaches are reported in **Table 4**, together with experimental results for the cubic structure for comparison. The agreement between experimental and calculated values is rather good for the cubic phase. On the other hand, values for the hexagonal structure seems to be overestimated in the calculated results compared to present experiments. In fact, they show values similar to those obtained for **s3** and **s4** samples, which however are related to nearly half-substituted solid solutions. Lower cell parameters values are expected for a fully Br-substituted hexagonal structure, as suggested by extrapolations to pure LiBr of fitted data (**Figure S10**), which give $a = 4.12$ Å and $c = 6.59$ Å. The comparison of calculated values for the orthorhombic structure with experimental ones is not possible, since no clear dissolution of Br into orthorhombic LiBH₄ was found experimentally.

### Assessment of the Phase Diagram

Assessed parameters of thermodynamic functions for different solution phases (hexagonal, cubic, orthorhombic and liquid) have been determined in the present study, in order to explore and characterize the $LiBH_4$–LiBr phase diagrams. In addition, pure components end-members have been also assessed. In both cases, the assessment procedure was based on results of experiments (**Table 5**) and of *ab initio* calculations (**Table 3**).

The literature reports only a single report supporting the solubility of LiBr into $o$-$LiBH_4$ up to 0.09 molar fraction.[14] However, this experimental result has been briefly describe in ref.[14] and it was not supported by detailed XRD analysis, thus it has not been considered for the assessment. On the other hand, no solubility of $LiBH_4$ into cubic LiBr has been observed (**Figure 1**).[12,14] For this reasons, positive parameters in the excess Gibbs free energy function for the orthorhombic and cubic phases have been fixed in the frame of the regular solution model, *i.e.* $^{ORTHO}\Omega = +3425$ J/mol and $^{CUB}\Omega = +20000$ J/mol, respectively. Considering the experimental value of enthalpy of mixing (*i.e.* -1.0±0.2 kJ/mol) for the sample **s5** (0.4LiBr-0.6LiBH₄ molar fraction), the interaction parameter for the hexagonal solid solution has been also fixed on the basis of a regular solution model and turns out to be $^{HEX}\Omega = -4167$ J/mol. In a first step, the liquid phase has been considered ideal and then it has been assessed based on liquidus and solidus temperatures taken from the literature[18] and obtained by DSC analysis in this work, as reported in **Figure S15** and **Table S2**. The optimised interaction parameter for the liquid phase assumed a negative value, $^{LIQ}\Omega = -2000$ J/mol.

| Phase | Experimental data | Calculated data |
|---|---|---|
| **ORTHO** | $0 \leq X_{LiBr} \leq 0.09$ molar fraction at 30 °C (Ref.[14]) | $0 \leq X_{LiBr} \leq 0.02$ molar fraction at 30 °C |
| **HEXA** | $0.30 \leq X_{LiBr} \leq 0.55$ at 30 °C (TW) <br> $0.29 \leq X_{LiBr} \leq 0.50$ at 30 °C (Ref.[14]) | $0.30 \leq X_{LiBr} \leq 0.56$ at 30 °C |
| **CUBIC** | $X_{LiBr} = 1$ at 30 °C and 30°C (Ref.[14]) | $X_{LiBr} = 1$ at 30 °C |
| **LIQUID** <br> **s4**, $X_{LiBr} = 0.30$ <br> **s3**, $X_{LiBr} = 0.40$ <br> **s2**, $X_{LiBr} = 0.50$ <br> **s2**, $X_{LiBr} = 0.50$ | $T_{SOL-LIQ}$ = 330-361 °C, $\Delta H_m$ = 10 kJ/mol (TW) <br> $T_{SOL-LIQ}$ = 352-371 °C, $\Delta H_m$ = 12 kJ/mol (TW) <br> $T_{SOL-LIQ}$ = 366-373 °C, $\Delta H_m$ = 10 kJ/mol (TW) <br> $T_{LIQ}$ = 377.9 °C (Ref.[18]) | $T_{SOL-LIQ}$ = 341-357 °C, $\Delta H_m$ = 13.2 kJ/mol <br> $T_{SOL-LIQ}$ = 357-368 °C, $\Delta H_m$ = 13.4 kJ/mol <br> $T_{SOL-LIQ}$ = 370-377 °C, $\Delta H_m$ = 13.8 kJ/mol |

***Table 5.*** *Literature, experimental and calculated thermodynamic data in the LiBH₄-LiBr system. $\Delta H_m$ is the enthalpy of melting. TW = This work. Literature solubility limit from ref.[14] at $X_{LiBr}$ = 0.09 has not been considered for the assessment.*

Starting from results obtained from *ab initio* calculations, the orthorhombic and hexagonal LiBr end-members have been also assessed. In order to take into account the results of the in-situ XRD analysis as a function of temperature, the LiBr hexagonal end-member has been described introducing a temperature dependent parameter (**Table 6**). This was necessary to obtain a limited Br⁻ solubility at high temperature, which turns out as high as x(LiBr) = 0.58 at 207 °C (compared to the experimental result equal to 0.55 at 207 °C, see **Figure S4**).

| Assessed excess Gibbs free energy (J mol⁻¹) |
|---|
| $^{CUB}G(LiBH_4) = {^{ORTH}}G(LiBH_4) + 3600$ [ref.[16]] |
| $^{ORTHO}G(LiBr) = {^{CUB}}G(LiBr) + 4304$ |
| $^{HEX}G(LiBr) = {^{CUB}}G(LiBr) + 1325 + 3.2*T$ |

***Table 6.*** *Assessed Gibbs free energy of end-members in the LiBH₄-LiBr system. Energies are reported as J/mol of compound.*

Compared to *ab initio* results for end-members energies at 0 K, the CALPHAD optimised values, which are determined from data above room temperature, are closer to CRYSTAL calculations.

Calculated values from VASP and QE show slightly higher differences, though these differences are rather limited (*i.e.* of the order of 2-3 kJ/mol).

**Figure 5** shows the calculated LiBH₄–LiBr phase diagram as a result of the CALPHAD assessment, compared with literature and new experimental data determined in this work. Corresponding calculated data are reported in **Table 5**.

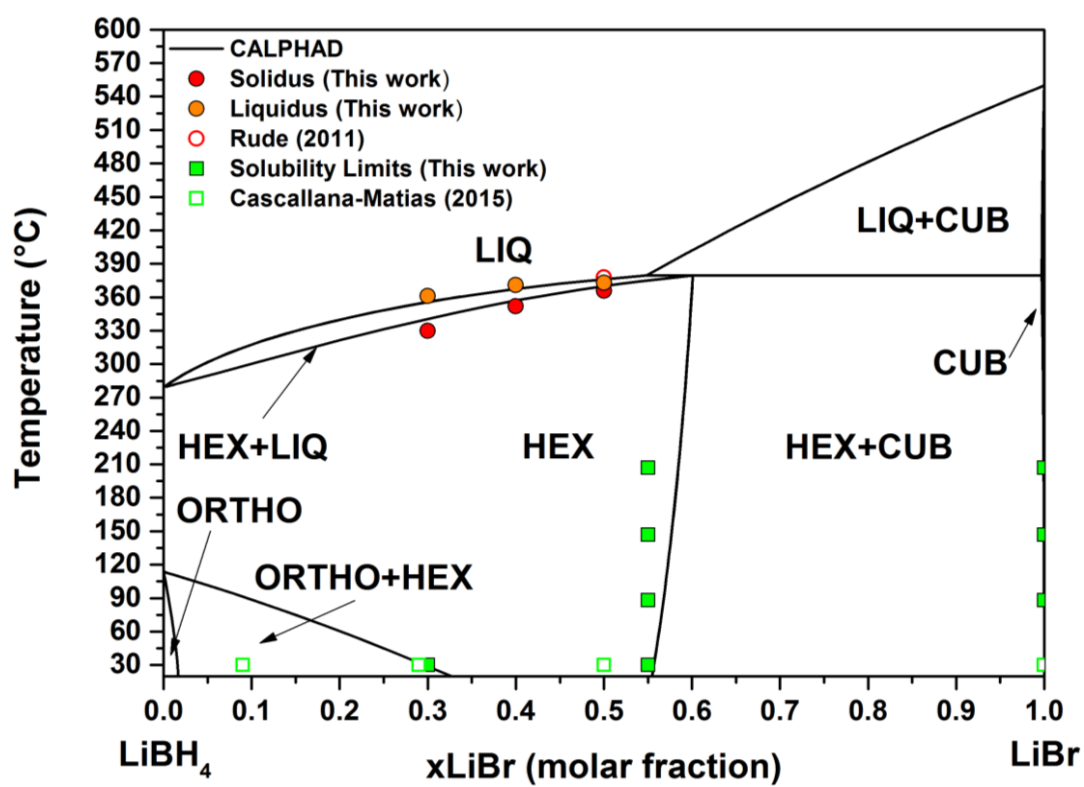

***Figure 5.*** *Lines: assessed phase diagram for the LiBH₄–LiBr system. Orange circles: experimental liquidus data from this work; red-open circle: liquidus temperature from ref.[18]; red circles: experimental liquidus data from this work; green squares: experimental solubility limits from this work; green-open square: literature solubility limit from ref.[14].*

It is worth noting that the calculated solubility limits of the hexagonal phase (both with orthorhombic and cubic phases) are in good agreement with experimental values at room temperature (*i.e.* 30 °C) determined by this study. If a solubility of LiBr in *o*-LiBH₄ up to 0.09 molar fraction[14] would be taken into account, a $^{\text{ORTHO}}\Omega$ = -1000 J/mol should be considered. In this case, the solubility limit of the hexagonal phase with the orthorhombic solid solution at room temperature would increase accordingly (**Figure S15**), deviating from experimental findings. So, as stated above, the experimental point at $X_{LiBr} = 0.09$ has not been considered for the assessment. On the other hand, the calculated solvus line of the hexagonal phase in the center of the diagram shows a slight deviation at high temperatures, compared to present experimental investigation, which essentially establishes a straight line as a function of temperature. The formation of a peritectic reaction at 380 °C and 0.60 LiBr molar fraction has been evidenced. The assessment provided a solubility up to 0.02 molar fraction of LiBr into o-LiBH₄ (**Table 5**). The calculated CALPHAD values of the liquidus temperature for samples **s2**, **s3** and **s4** (**Figure S16**) result in good agreement with respect to the experimental values (**Table 5**), as well as for melting enthalpy values upon heating (**Table 5** and **Table S2**). The average melting enthalpy for **s4** is lower with respect to **s2**, **s3**, and the calculated value, which could be related to a possible incipient decomposition of the sample due to the high temperatures reached in order to observe complete melting. Furthermore, lower experimental enthalpy values are detected during cooling, possibly because of a larger temperature range in which the crystallisation takes place hence causing an underestimation of the enthalpy related to the transformation while integrating the DSC peaks.

## Conclusions

The LiBH₄-LiBr phase diagram has been explored experimentally by means of PXD, *in situ* SR-PXD and DSC and thermodynamically assessed using the CALPHAD method coupled with *ab initio* results. The monophasic zone of the hexagonal Li(BH₄)₁₋ₓ(Br)ₓ solid solution has been defined equal to $0.30 \leq x \leq 0.55$ at R*T*. Various points of the phase diagram, including liquidus, solidus and solvus temperatures, have been investigated experimentally. In order to perform the assessment of the phase diagram, a value of the enthalpy of mixing has been estimated experimentally. A hand mixed sample (0.6LiBH₄-0.4LiBr) has been analysed by DSC in order to detect the exothermic peak due to the reaction for the formation of the hexagonal solid solution. Using the $\Delta H_{Exo}$ obtained by the calorimetric analysis, it was possible to conclude that the stabilization of the hexagonal solid solution starts after the LiBH$_4$ phase transition (*i.e.* at about 110 °C). Finally, the combination of DSC and PXD analyses allowed to define a $\Delta H_{Mix}$ equal to -1.0 ±0.2 kJ/mol for the complete formation of the Li(BH$_4$)$_{0.6}$(Br)$_{0.4}$ solid solution. Positive interaction parameters have been assessed for the orthorhombic and cubic phases, resulting in a limited solubility of LiBr into LiBH$_4$ orthorhombic structure and of LiBH$_4$ into LiBr cubic structure. On the contrary, the hexagonal phase has an extended stability area, so that an interaction parameter equal to $^{HEX}\Omega$ = -4167 J/mol has been considered. The liquid phase has been assessed based on literature and new experimental data and an interaction parameter equal to $^{LIQ}\Omega$ = -2000 J/mol has been obtained, resulting in a peritectic reaction at 380 °C and 0.60 LiBr molar fraction. The CALPHAD calculated transformation lines showed a good accordance with available literature and present experimental data.

## Supplementary Information

R$T$ PXD, IR-ATR and *in situ* SR-PXD patterns for various samples at different temperatures. Lattice parameters as a function of bromide content. DSC traces and solubilisation reaction followed by PXD. DSC traces and data ($\Delta H$ and temperatures) collected during the analysis.

## Acknowledgement

MC acknowledges Stefano Pantaleone and Piero Ugliengo, University of Torino, for fruitful discussion and the CRYSTAL team for continuous support in the usage of the CRYSTAL program. MP would like to thank the C3S supercomputing facility for the support with the OCCAM SuperComputer. VG acknowledges Asya Mazzucco and Gianluca Fiore for their support in DSC and XRD experiments. Part of this research was carried out at PETRA III beamline P02.1 at DESY synchrotron in Hamburg, Germany, a member of the Helmholtz Association (HGF). We would like to thank Alexander Schoekel, Claudio Pistidda and Christian Horstmann for assistance during the experiment "I-20180424 EC". Beamtime travel cost has been supported by the project CALIPSOplus under the Grant Agreeno

# References

(1)     Matsuo, M.; Orimo, S. Lithium Fast-Ionic Conduction in Complex Hydrides: Review and Prospects. *Adv. Energy Mater.* **2011**, *1* (2), 161–172. https://doi.org/10.1002/aenm.201000012.

(2)     Soulie, J.; Renaudin, G.; Černý, R.; Yvon, K. Lithium Boro-Hydride LiBH$_4$ I. Crystal Structure. *J. Alloys Compd.* **2002**, *346*, 200–205. https://doi.org/https://doi.org/10.1016/S0925-8388(02)00521-2.

(3)     El Kharbachi, A.; Pinatel, E.; Nuta, I.; Baricco, M. A Thermodynamic Assessment of LiBH$_4$. *Calphad* **2012**, *39*, 80–90. https://doi.org/10.1016/j.calphad.2012.08.005.

(4)     Matsuo, M.; Nakamori, Y.; Orimo, S.; Maekawa, H.; Takamura, H. Lithium Superionic Conduction in Lithium Borohydride Accompanied by Structural Transition. *Appl. Phys. Lett.* **2007**, *91* (22), 224103. https://doi.org/10.1063/1.2817934.

(5)     Gulino, V.; Wolczyk, A.; Golov, A. A.; Eremin, R. A.; Palumbo, M.; Nervi, C.; Blatov, V. A.; Proserpio, D. M.; Baricco, M. Combined DFT and Geometrical–Topological Analysis of Li-Ion Conductivity in Complex Hydrides. *Inorg. Chem. Front.* **2020**, *7* (17), 3115–3125. https://doi.org/10.1039/D0QI00577K.

(6)     Choi, Y. S.; Lee, Y.-S.; Oh, K. H.; Cho, Y. W. Interface-Enhanced Li Ion Conduction in a LiBH$_4$–SiO$_2$ Solid Electrolyte. *Phys. Chem. Chem. Phys.* **2016**, *18* (32), 22540–22547. https://doi.org/10.1039/C6CP03563A.

(7)     Blanchard, D.; Nale, A.; Sveinbjörnsson, D.; Eggenhuisen, T. M.; Verkuijlen, M. H. W.; Suwarno; Vegge, T.; Kentgens, A. P. M.; de Jongh, P. E. Nanoconfined LiBH$_4$ as a Fast Lithium Ion Conductor. *Adv. Funct. Mater.* **2015**, *25* (2), 184–192. https://doi.org/10.1002/adfm.201402538.

(8)     Gulino, V.; Barberis, L.; Ngene, P.; Baricco, M.; de Jongh, P. E. Enhancing Li-Ion Conductivity in LiBH$_4$-Based Solid Electrolytes by Adding Various Nanosized Oxides. *ACS Appl. Energy Mater.* **2020**, *3* (5), 4941–4948. https://doi.org/10.1021/acsaem.9b02268.

(9)     Gulino, V.; Brighi, M.; Murgia, F.; Ngene, P.; de Jongh, P.; Černý, R.; Baricco, M. Room-Temperature Solid-State Lithium-Ion Battery Using a LiBH$_4$–MgO Composite Electrolyte. *ACS Appl. Energy Mater.* **2021**, *4* (2), 1228–1236. https://doi.org/10.1021/acsaem.0c02525.

(10)    Maekawa, H.; Matsuo, M.; Takamura, H.; Ando, M.; Noda, Y.; Karahashi, T.; Orimo, S. Halide-Stabilized LiBH$_4$, a Room-Temperature Lithium Fast-Ion Conductor. *J. Am. Chem. Soc.* **2009**, *131* (3), 894–895. https://doi.org/10.1021/ja807392k.

(11)    Oguchi, H.; Matsuo, M.; Hummelshøj, J. S.; Vegge, T.; Nørskov, J. K.; Sato, T.; Miura, Y.; Takamura, H.; Maekawa, H.; Orimo, S. Experimental and Computational Studies on Structural

Transitions in the LiBH$_4$–LiI Pseudobinary System. *Appl. Phys. Lett.* **2009**, *94* (14), 141912. https://doi.org/10.1063/1.3117227.

(12)  Gulino, V.; Brighi, M.; Dematteis, E. M.; Murgia, F.; Nervi, C.; Černý, R.; Baricco, M. Phase Stability and Fast Ion Conductivity in the Hexagonal LiBH$_4$–LiBr–LiCl Solid Solution. *Chem. Mater.* **2019**, *31* (14), 5133–5144. https://doi.org/10.1021/acs.chemmater.9b01035.

(13)  I, R. M.; Karahashi, T.; Kumatani, N.; Noda, Y.; Ando, M.; Takamura, H.; Matsuo, M.; Orimo, S.; Maekawa, H. Room Temperature Lithium Fast-Ion Conduction and Phase Relationship of LiI Stabilized LiBH$_4$. *Solid State Ionics* **2011**, *192* (1), 143–147. https://doi.org/10.1016/j.ssi.2010.05.017.

(14)  Cascallana-Matias, I.; Keen, D. A.; Cussen, E. J.; Gregory, D. H. Phase Behavior in the LiBH$_4$– LiBr System and Structure of the Anion-Stabilized Fast Ionic, High Temperature Phase. *Chem. Mater.* **2015**, *27* (22), 7780–7787. https://doi.org/10.1021/acs.chemmater.5b03642.

(15)  Lukas, H.; Fries, S. G.; Sundman, B. *Computational Thermodynamics*; Cambridge University Press: Cambridge, 2007. https://doi.org/10.1017/CBO9780511804137.

(16)  Dematteis, E. M.; Roedern, E.; Pinatel, E. R.; Corno, M.; Jensen, T. R.; Baricco, M. A Thermodynamic Investigation of the LiBH$_4$–NaBH$_4$ System. *RSC Adv.* **2016**, *6* (65), 60101– 60108. https://doi.org/10.1039/C6RA09301A.

(17)  Dematteis, E. M.; Pinatel, E. R.; Corno, M.; Jensen, T. R.; Baricco, M. Phase Diagrams of the LiBH$_4$–NaBH$_4$–KBH$_4$ System. *Phys. Chem. Chem. Phys.* **2017**, *19* (36), 25071–25079. https://doi.org/10.1039/C7CP03816J.

(18)  Rude, L. H.; Zavorotynska, O.; Arnbjerg, L. M.; Ravnsbæk, D. B.; Malmkjær, R. A.; Grove, H.; Hauback, B. C.; Baricco, M.; Filinchuk, Y.; Besenbacher, F.; Jensen, T. R. Bromide Substitution in Lithium Borohydride, LiBH$_4$–LiBr. *Int. J. Hydrogen Energy* **2011**, *36* (24), 15664–15672. https://doi.org/10.1016/j.ijhydene.2011.08.087.

(19)  Lutterotti, L., Matthies, S., Wenk, H. R. MAUD: A Friendly Java Program for Material Analysis Using Diffraction. *MAUD A Friendly Java Progr. Mater. Anal. Using Diffr.* **1999**, 14–15.

(20)  Perdew, J. P.; Burke, K.; Ernzerhof, M. Generalized Gradient Approximation Made Simple. *Phys. Rev. Lett.* **1996**, *77* (18), 3865–3868. https://doi.org/10.1103/PhysRevLett.77.3865.

(21)  Dovesi, R.; Saunders, V. R.; Roetti, C.; Orlando, R.; Zicovich-Wilson, C. M.; Pascale, F.; Civalleri, B.; Doll, K.; Harrison, N. M.; Bush, I. J.; D'Arco, P.; Llunell, M.; Causà, M.; Noël, Y. *CRYSTAL14 User's Manual, University of Torino*; 2014.

(22)  Dovesi, R.; Erba, A.; Orlando, R.; Zicovich-Wilson, C. M.; Civalleri, B.; Maschio, L.; Rérat, M.; Casassa, S.; Baima, J.; Salustro, S.; Kirtman, B. Quantum-Mechanical Condensed Matter

Simulations with CRYSTAL. *Wiley Interdiscip. Rev. Comput. Mol. Sci.* **2018**, *8* (4), e1360. https://doi.org/10.1002/wcms.1360.

(23) Doll, K.; Stoll, H. Ground-State Properties of Heavy Alkali Halides. *Phys. Rev. B* **1998**, *57* (8), 4327–4331. https://doi.org/10.1103/PhysRevB.57.4327.

(24) Pascale, F.; Zicovich-Wilson, C. M.; López Gejo, F.; Civalleri, B.; Orlando, R.; Dovesi, R. The Calculation of the Vibrational Frequencies of Crystalline Compounds and Its Implementation in the CRYSTAL Code. *J. Comput. Chem.* **2004**, *25* (6), 888–897. https://doi.org/10.1002/jcc.20019.

(25) Zicovich-Wilson, C. M.; Torres, F. J.; Pascale, F.; Valenzano, L.; Orlando, R.; Dovesi, R. Ab Initio Simulation of the IR Spectra of Pyrope, Grossular, and Andradite. *J. Comput. Chem.* **2008**, *29* (13), 2268–2278. https://doi.org/10.1002/jcc.20993.

(26) Kresse, G.; Hafner, J. Ab Initio Molecular Dynamics for Liquid Metals. *Phys. Rev. B* **1993**, *47* (1), 558–561. https://doi.org/10.1103/PhysRevB.47.558.

(27) Kresse, G.; Furthmüller, J. Efficiency of Ab-Initio Total Energy Calculations for Metals and Semiconductors Using a Plane-Wave Basis Set. *Comput. Mater. Sci.* **1996**, *6* (1), 15–50. https://doi.org/10.1016/0927-0256(96)00008-0.

(28) Kresse, G.; Furthmüller, J. Efficient Iterative Schemes for Ab Initio Total-Energy Calculations Using a Plane-Wave Basis Set. *Phys. Rev. B* **1996**, *54* (16), 11169–11186. https://doi.org/10.1103/PhysRevB.54.11169.

(29) Monkhorst, H. J.; Pack, J. D. Special Points for Brillouin-Zone Integrations. *Phys. Rev. B* **1976**, *13* (12), 5188–5192. https://doi.org/10.1103/PhysRevB.13.5188.

(30) Methfessel, M.; Paxton, A. T. High-Precision Sampling for Brillouin-Zone Integration in Metals. *Phys. Rev. B* **1989**, *40* (6), 3616–3621. https://doi.org/10.1103/PhysRevB.40.3616.

(31) Blöchl, P. E.; Jepsen, O.; Andersen, O. K. Improved Tetrahedron Method for Brillouin-Zone Integrations. *Phys. Rev. B* **1994**, *49* (23), 16223–16233. https://doi.org/10.1103/PhysRevB.49.16223.

(32) Togo, A.; Tanaka, I. First Principles Phonon Calculations in Materials Science. *Scr. Mater.* **2015**, *108*, 1–5. https://doi.org/10.1016/j.scriptamat.2015.07.021.

(33) Giannozzi, P.; Andreussi, O.; Brumme, T.; Bunau, O.; Buongiorno Nardelli, M.; Calandra, M.; Car, R.; Cavazzoni, C.; Ceresoli, D.; Cococcioni, M.; Colonna, N.; Carnimeo, I.; Dal Corso, A.; de Gironcoli, S.; Delugas, P.; DiStasio, R. A.; Ferretti, A.; Floris, A.; Fratesi, G.; Fugallo, G.; Gebauer, R.; Gerstmann, U.; Giustino, F.; Gorni, T.; Jia, J.; Kawamura, M.; Ko, H.-Y.; Kokalj, A.; Küçükbenli, E.; Lazzeri, M.; Marsili, M.; Marzari, N.; Mauri, F.; Nguyen, N. L.; Nguyen, H.-V.; Otero-de-la-Roza, A.; Paulatto, L.; Poncé, S.; Rocca, D.; Sabatini, R.; Santra,

B.; Schlipf, M.; Seitsonen, A. P.; Smogunov, A.; Timrov, I.; Thonhauser, T.; Umari, P.; Vast, N.; Wu, X.; Baroni, S. Advanced Capabilities for Materials Modelling with Quantum ESPRESSO. *J. Phys. Condens. Matter* **2017**, *29* (46), 465901. https://doi.org/10.1088/1361-648X/aa8f79.

(34) Giannozzi, P.; Baroni, S.; Bonini, N.; Calandra, M.; Car, R.; Cavazzoni, C.; Ceresoli, D.; Chiarotti, G. L.; Cococcioni, M.; Dabo, I.; Dal Corso, A.; de Gironcoli, S.; Fabris, S.; Fratesi, G.; Gebauer, R.; Gerstmann, U.; Gougoussis, C.; Kokalj, A.; Lazzeri, M.; Martin-Samos, L.; Marzari, N.; Mauri, F.; Mazzarello, R.; Paolini, S.; Pasquarello, A.; Paulatto, L.; Sbraccia, C.; Scandolo, S.; Sclauzero, G.; Seitsonen, A. P.; Smogunov, A.; Umari, P.; Wentzcovitch, R. M. QUANTUM ESPRESSO: A Modular and Open-Source Software Project for Quantum Simulations of Materials. *J. Phys. Condens. Matter* **2009**, *21* (39), 395502. https://doi.org/10.1088/0953-8984/21/39/395502.

(35) www.quantum-espresso.org.

(36) Dal Corso, A. Pseudopotentials Periodic Table: From H to Pu. *Comput. Mater. Sci.* **2014**, *95*, 337–350. https://doi.org/10.1016/j.commatsci.2014.07.043.

(37) Kister, A. T.; Redlich, O. Algebraic Representation of Thermodynamic Properties and the Classification. *Ind. Eng. Chem.* **1948**, *40* (2), 345–348.

(38) Rapp, J. E.; Merchant, H. D. Thermal Expansion of Alkali Halides from 70 to 570 K. *J. Appl. Phys.* **1973**, *44* (9), 3919–3923. https://doi.org/10.1063/1.1662872.

(39) Filinchuk, Y.; Chernyshov, D.; Cerny, R. Lightest Borohydride Probed by Synchrotron X-Ray Diffraction: Experiment Calls for a New Theoretical Revision. *J. Phys. Chem. C* **2008**, *112* (28), 10579–10584. https://doi.org/10.1021/jp8025623.

(40) Arnbjerg, L. M.; Ravnsbæk, D. B.; Filinchuk, Y.; Vang, R. T.; Cerenius, Y.; Besenbacher, F.; Jørgensen, J.-E.; Jakobsen, H. J.; Jensen, T. R. Structure and Dynamics for LiBH₄–LiCl Solid Solutions. *Chem. Mater.* **2009**, *21* (24), 5772–5782. https://doi.org/10.1021/cm902013k.

(41) Zavorotynska, O.; Corno, M.; Pinatel, E.; Rude, L. H.; Ugliengo, P.; Jensen, T. R.; Baricco, M. Theoretical and Experimental Study of LiBH₄-LiCl Solid Solution. *Crystals* **2012**, *2* (4), 144–158. https://doi.org/10.3390/cryst2010144.

(42) Hagemann, H.; Gomes, S.; Renaudin, G.; Yvon, K. Raman Studies of Reorientation Motions of [BH₄]⁻ Anionsin Alkali Borohydrides. *J. Alloys Compd.* **2004**, *363* (1–2), 129–132. https://doi.org/10.1016/S0925-8388(03)00468-7.

(43) Hagemann, H.; Filinchuk, Y.; Chernyshov, D.; Van Beek, W. Lattice Anharmonicity and Structural Evolution of LiBH₄: An Insight from Raman and X-Ray Diffraction Experiments. *Phase Transitions* **2009**, *82* (4), 344–355. https://doi.org/10.1080/01411590802707688.

(44)  Rude, L. H.; Groppo, E.; Arnbjerg, L. M.; Ravnsbæk, D. B.; Malmkjær, R. A.; Filinchuk, Y.; Baricco, M.; Besenbacher, F.; Jensen, T. R. Iodide Substitution in Lithium Borohydride, LiBH$_4$–LiI. *J. Alloys Compd.* **2011**, *509* (33), 8299–8305. https://doi.org/10.1016/j.jallcom.2011.05.031.

(45)  El Kharbachi, A.; Nuta, I.; Hodaj, F.; Baricco, M. Above Room Temperature Heat Capacity and Phase Transition of Lithium Tetrahydroborate. *Thermochim. Acta* **2011**, *520* (1–2), 75–79. https://doi.org/10.1016/j.tca.2011.02.043.

(46)  Čančarević, Ž. P.; Schön, J. C.; Jansen, M. Stability of Alkali Metal Halide Polymorphs as a Function of Pressure. *Chem. - An Asian J.* **2008**, *3* (3), 561–572. https://doi.org/10.1002/asia.200700323.

(47)  Sure, R.; Grimme, S. Corrected Small Basis Set Hartree-Fock Method for Large Systems. *J. Comput. Chem.* **2013**, *34* (19), 1672–1685. https://doi.org/10.1002/jcc.23317.

(48)  Grimme, S.; Antony, J.; Ehrlich, S.; Krieg, H. A Consistent and Accurate Ab Initio Parametrization of Density Functional Dispersion Correction (DFT-D) for the 94 Elements H-Pu. *J. Chem. Phys.* **2010**, *132* (15). https://doi.org/10.1063/1.3382344.

(49)  Cortona, P. Direct Determination of Self-Consistent Total Energies and Charge Densities of Solids: A Study of the Cohesive Properties of the Alkali Halides. *Phys. Rev. B* **1992**, *46* (4), 2008–2014. https://doi.org/10.1103/PhysRevB.46.2008.